# Supermembrane with Non-Abelian Gauging and Chern - Simons Quantization[1]

Hitoshi NISHINO[2] and Subhash RAJPOOT[3]

*Department of Physics & Astronomy*
*California State University*
*1250 Bellflower Boulevard*
*Long Beach, CA 90840*

## Abstract

We present non-Abelian gaugings of supermembrane for general isometries for compactifications from eleven-dimensions, starting with Abelian case as a guide. We introduce a super Killing vector in eleven-dimensional superspace for a non-Abelian group $G$ associated with the compact space $B$ for a general compactification, and couple it to a non-Abelian gauge field on the world-volume. As a technical tool, we use teleparallel superspace with no manifest local Lorentz covariance. Interestingly, the coupling constant is quantized for the non-Abelian group $G$, due to its generally non-trivial mapping $\pi_3(G)$.

---

[1]Work supported in part by NSF Grant # 0308246
[2]E-Mail: hnishino@csulb.edu
[3]E-Mail: rajpoot@csulb.edu

## 1. Introduction

The concept of the simultaneous double-compactification of supermembrane on three-dimensions (3d) with target eleven-dimensions (11D) into superstring on 2d with target 10D, was first presented in [1]. Since this first observation, it has been well-known that massive Type IIA supergravity in 10D [2] can also arise from the compactification of M-theory in 11D [3], *via* a Killing vector in the direction of the compactifying 11-th coordinate [4]. This mechanism has been elucidated in terms of component language [4]. Similar mechanisms are expected to work also in many other dimensional reductions [5].

At the present time, however, it is not clear how these component results can be reformulated in 11D superspace [6][7] with symmetries for supermembrane action [8]. For example, the original important significance of supermembrane, such as fermionic $\kappa$-invariance [9][8], or target 11D superspace Bianchi identities (BIds) [6][7], has not been clarified in component language [4]. Neither is it clear in [4] how such a theory as 'unique' as 11D supergravity [3] can accommodate the 'free' mass parameter $m$, or how it makes itself equivalent to the conventional theory [3], while generating massive Type IIA supergravity in 10D [2] after the compactification.

In this paper, we will clarify the significance of the 'free' parameter $m$ in the context of supermembrane [8] on 11D superspace background [7]. We first review the modification of 11D supergravity with the modified fourth-rank field strength by a Killing vector with the free parameter $m$ [4] in component language. We see that all the $m$-terms cancel themselves in Bianchi identities, when the field strength is expressed in terms of Lorentz indices. We next show how such a disappearance of $m$-effects is reformulated in superspace [6][7] as well. In other words, there is no effect by the $m$-dependent terms in superspace, with no significance or physical effects by $m$-modifications.

At first glance, this result seems discouraging, because any effect by super Killing vector corresponding to the compactification from 11D into 10D turns out to be 'phantom'. Interestingly, however, we have also found that if we introduce an $U(1)$ gauge field on the supermembrane world-volume with a minimal coupling to a super Killing vector $\xi^A$, there surely is physical effect depending on $m$. We have also found that such couplings necessitate the existence of a Chern-Simons term. We can further generalize this $U(1)$ gauge group for a torus compactification into 10D, to a more general compactification with a more general non-Abelian isometry group. Fortunately, all the $m$-dependent terms do not upset the basic structure of supermembrane action.

Accordingly, the super Killing vector $\xi^{AI}$ for a non-Abelian group $G$ carries the adjoint index $I = 1, 2, \cdots, \dim G$, where $G$ is associated with the compact space $B$ in the compactification $M_{11} \to M_D \times B$ from 11D into any arbitrary space-time dimension $M_D$ with $D \equiv 11 - \dim B$ [10]. Typical examples are such as $G = SO(8)$ for $B = S^7$, or $G = SO(6) \times SO(3)$ for $B = S^5 \times S^2$. The simplest choice $G = SO(2)$ [4] is for the torus compactification $B = S^1$. In the series of generalized Scherk-Schwarz type [11] dimensional reductions $B = SL(11 - D, \mathbb{R})/SO(11 - D)$ [12][5] we have $G = SO(11 - D)$.

As a technical tool, we use a special set of 11D superspace constraints named teleparallel superspace constraints [13]. This is because compactifications from 11D most naturally break local Lorentz symmetry, and therefore, teleparallel superspace with no manifest local Lorentz symmetry is more suitable for such a formulation.

Our vector field on the world sheet is neither auxiliary nor composite, but is topological, and different from the auxiliary vector field introduced in massive type IIA formulation [4]. It is also distinct from the $U(1)$ vector field used in D-brane formulation [14], even though we leave the possibility of an important connection with the latter, for future studies.

## 2. Modified Field Strengths in Component

In this section, we study the effect of the Killing vector $\xi^m$ for the massive branes described in [4] on 11D supergravity in component language. The Killing vector $\xi^m$ is associated with the compactification of 11D supergravity [3] down to 10D massive Type IIA supergravity [2]. We claim that the additional $m$-dependent terms in a fourth-rank field strength [4] with $\xi^m$ can eventually disappear in its Bianchi identity, when the field strength is expressed with Lorentz indices.

The fourth-rank field strength $G_{mnrs}$ [3] of the potential $B_{mnr}$ is [4],[4]

$$\check{G}_{mnrs} \equiv \tfrac{1}{6}\partial_{[m}B_{nrs]} - \tfrac{1}{8}m\widetilde{B}_{[mn}\widetilde{B}_{rs]} \quad . \tag{2.1}$$

Here $\widetilde{B}_{mn} \equiv \xi^r B_{rmn}$ and $\widetilde{\Lambda}_m \equiv \xi^n \Lambda_{nm}$. More generally, any *tilded* field or parameter implies a contraction with $\xi^m$ from the left corresponding to the $i_\xi$-operation in terms of differential forms [4]. The Killing vector $\xi^m$ specifies the 11-th direction of the compactification [4], associated with the Lie-derivatives

$$\mathcal{L}_\xi B_{mnr} \equiv \xi^s \partial_s B_{mnr} + \tfrac{1}{2}(\partial_{[m|}\xi^s)B_{s|nr]} \stackrel{*}{=} 0 \quad , \tag{2.2a}$$

$$\mathcal{L}_\xi g_{mn} \equiv \xi^r \partial_r g_{mn} + (\partial_{(m|}\xi^r)g_{r|n)} \stackrel{*}{=} 0 \quad , \tag{2.2b}$$

$$\mathcal{L}_\xi e_m{}^a \equiv \xi^n \partial_n e_m{}^a + (\partial_m \xi^n)e_n{}^a \stackrel{*}{=} 0 \quad , \tag{2.2c}$$

---

[4]Our notation for the curved (or Lorentz) indices $m, n, \cdots$ (or $a, b, \cdots$) are the same as in [6]. Also our antisymmetrization is as in [6], *e.g.*, $A_{[m}B_{n]} \equiv A_m B_n - B_n A_m$ with *no* 1/2 in front.

$$E_a \xi^b \stackrel{*}{=} \xi^c C_{ca}{}^b \ , \tag{2.2d}$$

$$\mathcal{L}_\xi C_{ab}{}^c = \xi^d E_d C_{ab}{}^c \stackrel{*}{=} 0 \ , \tag{2.2e}$$

where $E_a \equiv e_a{}^m \partial_m$ and $C_{ab}{}^c$ is the anholonomy coefficient $C_{ab}{}^c \equiv (E_{[a} e_{b]}{}^m) e_m{}^c$ both with *no* Lorentz connection, because we are in teleparallel formulation. The symbol $\stackrel{*}{=}$ stands for a relationship associated with the feature of the Killing vector. As we will also see, our engagement of teleparallel formulation is compatible with the Killing vector condition. Eq. (2.2e) can be easily confirmed by (2.2d). As far as the target 11D superspace is concerned, there will be no physical difference between teleparallel formulation [13] and the conventional one [7], as has been explained also in [13].

The real meaning of the $m$-modification becomes clear, when we rewrite this field strength in terms of local Lorentz indices:

$$\check{G}_{abcd} \equiv +\tfrac{1}{6} E_{[a} B_{bcd]} - \tfrac{1}{4} \check{C}_{[ab|}{}^e B_{e|cd]} + \tfrac{1}{8} m \widetilde{B}_{[ab} \widetilde{B}_{cd]}$$
$$= G_{abcd} - \tfrac{1}{8} m \widetilde{B}_{[ab} \widetilde{B}_{cd]} \ , \tag{2.3a}$$

$$G_{abcd} \equiv +\tfrac{1}{6} E_{[a} B_{bcd]} - \tfrac{1}{4} C_{[ab|}{}^e B_{e|cd]} \ , \tag{2.3b}$$

where we have used also the modified anholonomy coefficients

$$\check{C}_{ab}{}^c = C_{ab}{}^c + m \widetilde{B}_{ab} \xi^c \ , \tag{2.4}$$

consistent with the torsion $T_{mn}{}^r = -m \widetilde{B}_{mn} \xi^r$ in [4]. Here $C_{ab}{}^c$ is the original anholonomy coefficient at $m = 0$ [3]. The 'disappearance' of the $m$-effect can be understood by the $\chi$-gauge transformation in [4] that we rename $\Lambda$-transformation here:

$$\delta_\Lambda B_{mnr} = \tfrac{1}{2} \partial_{[m} \Lambda_{nr]} - \tfrac{1}{2} m \widetilde{\Lambda}_{[m} \widetilde{B}_{nr]} \ . \tag{2.5}$$

This together with other related ones can be expressed mostly with Lorentz indices, as[5]

$$\delta_\Lambda B_{abc} = +\tfrac{1}{2} E_{[a} \Lambda_{bc]} - \tfrac{1}{2} \check{C}_{[ab|}{}^d \Lambda_{d|c]} + \tfrac{1}{2} m \widetilde{\Lambda}_{[a} \widetilde{B}_{bc]}$$
$$= +\tfrac{1}{2} E_{[a} \Lambda_{bc]} - \tfrac{1}{2} C_{[ab|}{}^d \Lambda_{d|c]}$$
$$= \delta_\Lambda B_{abc} \Big|_{m=0} \ , \tag{2.6a}$$

$$\delta_\Lambda e_a{}^m = +m \widetilde{\Lambda}_a \xi^m \ , \quad \delta_\Lambda e_m{}^a = -m \widetilde{\Lambda}_m \xi^a \ , \quad \delta_\Lambda g_{mn} = -m \widetilde{\Lambda}_{(m} \xi_{n)} \ , \tag{2.6b}$$

$$\delta_\Lambda \xi^m = 0 \ , \quad \delta_\Lambda \xi^a = 0 \ , \tag{2.6c}$$

$$\delta_\Lambda \check{G}_{mnrs} = +\tfrac{1}{6} m \widetilde{\Lambda}_{[m} \widetilde{G}_{nrs]} \ , \quad \widetilde{G}_{mnr} \equiv \xi^s \check{G}_{smnr} \ , \tag{2.6d}$$

$$\delta_\Lambda \check{G}_{abcd} = 0 \ . \tag{2.6e}$$

---

[5]The check-symbol on $\widetilde{G}_{abc}$ in (2.6d) is *not* needed, because $\xi^s \widetilde{G}_{smnr} \equiv \xi^s G_{smnr}$.

Most importantly, when written in terms of Lorentz indices, the field strength $\check{G}_{abcd}$ is neutral under the $\Lambda$-transformation. On the other hand, as (2.6d) shows in agreement with [4], $\check{G}_{mnrs}$ with curved indices is *not* invariant. The reason is that the elfbein transformation $\delta_\Lambda e_a{}^m$ cancels exactly the contribution of $\delta_\Lambda \check{G}_{mnrs}$. Relevantly, all the $m$-dependent terms in (2.6a) are completely cancelled by themselves, making the whole expression exactly the same as the $m=0$ case.

Relevantly, $C$, $G$, $\check{C}$ and $\check{G}$ satisfy the BIds in component language[6]

$$\tfrac{1}{2} E_{[a} C_{bc]}{}^d - \tfrac{1}{2} C_{[ab|}{}^e C_{e|c]}{}^d \equiv 0 \ , \tag{2.7a}$$

$$\tfrac{1}{24} E_{[a} G_{bcde]} - \tfrac{1}{12} C_{[ab|}{}^f G_{f|cde]} \equiv 0 \ , \tag{2.7b}$$

$$\tfrac{1}{2} E_{[a} \check{C}_{bc]}{}^d - \tfrac{1}{2} \check{C}_{[ab|}{}^e \check{C}_{e|c]}{}^d + m \widetilde{G}_{abc} \xi^d \equiv 0 \ , \tag{2.7c}$$

$$\tfrac{1}{24} E_{[a} \check{G}_{bcde]} - \tfrac{1}{12} \check{C}_{[ab|}{}^f \check{G}_{f|cde]} \equiv 0 \ . \tag{2.7d}$$

Eq. (2.7c) and (2.7d) are equivalent to (2.7a) and (2.7b), reflecting again the disappearance of the $m$-terms in (2.6a). To put it differently, we can confirm (2.7c) and (2.7d), using (2.7a) and (2.7b). In this process, we need the property that $\widetilde{G}_{abc}$ satisfies its 'own' BId[7]

$$\tfrac{1}{6} E_{[a} \widetilde{G}_{bcd]} - \tfrac{1}{4} C_{[ab|}{}^e \widetilde{G}_{e|cd]} \equiv 0 \ . \tag{2.8}$$

Relevantly, we can show that $\widetilde{G}_{abc}$ also equal

$$\widetilde{G}_{abc} \equiv \xi^d G_{dabc} = -\left( \tfrac{1}{2} E_{[a} \widetilde{B}_{bc]} - \tfrac{1}{2} \check{C}_{[ab|}{}^d \widetilde{B}_{d|c]} \right) \ . \tag{2.9}$$

The first equality is the original definition, while the second one can be confirmed by the use of (2.3b). The overall negative sign is due to our definition of *tilded* fields.

As has been mentioned before, (2.2d) has no Lorentz connection. The consistency of our teleparallelism is justified by the consistency of the commutator of the $E_a$'s on $\xi^c$. In fact, we get

$$[E_a, E_b] \xi^c = E_{[a}(E_{b]} \xi^c) = C_{ab}{}^d E_d \xi^c + \xi^d E_d C_{ab}{}^c \ , \tag{2.10}$$

where from the middle side to r.h.s., we have used (2.2d) and the BI (2.7a). As desired, the first term on the r.h.s. coincides with the l.h.s., while the last term vanishes, thanks to (2.2e).

---

[6]We note that in the earlier version of this paper, there was a redundant $\widetilde{G}\widetilde{B}$-term in the $\check{G}$-BId which should not have been there.

[7]The difference between $\check{C}_{ab}{}^e$ and $C_{ab}{}^e$ does not matter here, due to the identity $\xi^e \widetilde{G}_{ecd} \equiv 0$.

We have thus seen that all the $m$-dependent terms in the $\check{G}_{abcd}$-BId are cancelled, when this field strength is expressed with Lorentz indices. This means that all of these $m$-dependent terms do not really generate any new physical effect within 11D supergravity. This aspect will be used as the guiding principle in reformulation in superspace [6][7] in the next section. This result of no 'physical' effect of the Killing vector [4] in 11D supergravity [3] is not surprising. This is because 11D supergravity [3] is so unique and tight that there is no room for such a an additional free parameter $m$. We have adopted teleparallel formulation in component, but the necessity of this will be more elucidated in next sections, when the Killing vector is coupled to supermembrane.

## 3. Modified BIds in Superspace

We have seen that all the $m$-modified terms in the $\check{G}$-BId are completely absorbed into field redefinitions within 11D. We have shown this in terms of teleparallel formulation. This aspect should be reformulated in superspace [6][7], in particular in so-called teleparallel superspace developed in [13]. Let us start with the non-modified teleparallel superspace with the super anholonomy coefficients $C$- and superfield strength $G$ defined by[13][8]

$$C_{AB}{}^C \equiv (E_{[A} E_{B)}{}^M) E_M{}^C \ , \tag{3.1a}$$

$$G_{ABCD} \equiv \tfrac{1}{6} E_{[A} B_{BCD)} - \tfrac{1}{4} C_{[AB|}{}^E B_{E|CD)} \ , \tag{3.1b}$$

satisfying their BIds

$$\tfrac{1}{2} E_{[A} C_{BC)}{}^D - \tfrac{1}{2} C_{[AB|}{}^E C_{E|C)}{}^D \equiv 0 \ , \tag{3.1a}$$

$$\tfrac{1}{24} E_{[A} G_{BCDE)} - \tfrac{1}{12} C_{[AB|}{}^F G_{F|CDE)} \equiv 0 \ , \tag{3.1b}$$

where $E_A \equiv E_A{}^M \partial_M$ [6]. The superspace constraints at engineering dimensions $d \leq 1$ relevant at $m = 0$ are [13]

$$C_{\alpha\beta}{}^c = +i(\gamma^c)_{\alpha\beta} \ , \quad G_{\alpha\beta cd} = +\tfrac{1}{2}(\gamma_{cd})_{\alpha\beta} \ , \tag{3.2a}$$

$$C_{\alpha\beta}{}^\gamma = +\tfrac{1}{4}(\gamma_{de})_{(\alpha}{}^\gamma C_{\beta)}{}^{de} \ , \quad C_\alpha{}^{bc} = -C_\alpha{}^{cb} \ , \tag{3.2b}$$

$$C_{\alpha b}{}^\gamma = +\tfrac{i}{144}(\gamma_b{}^{defg} G_{defg} + 8\gamma^{def} G_{bdef})_\alpha{}^\gamma - \tfrac{1}{8}(\gamma^{cd})_\alpha{}^\gamma (2C_{bcd} - C_{cdb}) \ . \tag{3.2c}$$

All other independent components at $d \leq 1$ such as $G_{\alpha bcd}$ and $C_{\alpha\beta}{}^\gamma$ are all zero.

---

[8]As in [6], we use the indices $A, B, \cdots$ for local Lorentz coordinates in superspace, while $M, N, \cdots$ for curved ones.

The super Killing vector $\xi^M$ in superspace for Abelian gauging corresponds to the torus compactification $M_{11} \to M_{10} \times S^1$, specified by the conditions

$$\mathcal{L}_\xi B_{MNP} \equiv \xi^Q \partial_Q B_{MNP} + \tfrac{1}{2}(\partial_{[M|}\xi^Q) B_{Q|NP)} \overset{*}{=} 0 ~~, \tag{3.3a}$$

$$\mathcal{L}_\xi E_M{}^A \equiv \xi^N \partial_N E_M{}^A + (\partial_M \xi^N) E_N{}^A \overset{*}{=} 0 ~~, \tag{3.3b}$$

$$\mathcal{L}_\xi \xi^M \overset{*}{=} 0 ~~, \tag{3.3c}$$

$$E_A \xi^B \overset{*}{=} \xi^C C_{CA}{}^B ~~, \tag{3.3d}$$

$$\mathcal{L}_\xi C_{AB}{}^C = \xi^D E_D C_{AB}{}^C \overset{*}{=} 0 ~~. \tag{3.3e}$$

These are teleparallel superspace generalizations of the component case (2.2). Eq. (3.3d) is nothing but a rewriting of (3.3b). As in the component case (2.10), we can confirm the consistency of (3.3d) by considering the commutator $[E_A, E_B\}\xi^C$ with the aid of (3.3e), whose details are skipped here.

The BIds for the $m$-modified system with the Abelian super Killing vector are[9]

$$\tfrac{1}{2} E_{[A} \check{C}_{BC)}{}^D - \tfrac{1}{2} \check{C}_{[AB|}{}^E \check{C}_{E|C)}{}^D + m \widetilde{G}_{ABC} \xi^D \equiv 0 ~~, \tag{3.4a}$$

$$\tfrac{1}{24} E_{[A} \check{G}_{BCDE)} - \tfrac{1}{12} \check{C}_{[AB|}{}^F \check{G}_{F|CDE)} \equiv 0 ~~, \tag{3.4b}$$

$$\tfrac{1}{6} E_{[A} \widetilde{G}_{BCD)} - \tfrac{1}{4} \check{C}_{[AB|}{}^D \widetilde{G}_{D|CD)} \equiv 0 ~~, \tag{3.4c}$$

where the modified superfield strengths $\check{C}_{AB}{}^C$, $\check{G}_{ABCD}$ and $\widetilde{G}_{ABC}$ are defined by[10]

$$\check{C}_{AB}{}^C \equiv C_{AB}{}^C + m \widetilde{B}_{AB} \xi^C ~~, \tag{3.5a}$$

$$\check{G}_{ABCD} \equiv \tfrac{1}{6} E_{[A} B_{BCD)} - \tfrac{1}{4} \check{C}_{[AB|}{}^E B_{E|CD)} + \tfrac{1}{8} m \widetilde{B}_{[AB} \widetilde{B}_{CD)}$$

$$= G_{ABCD} - \tfrac{1}{8} m \widetilde{B}_{[AB} \widetilde{B}_{CD)} ~~, \tag{3.5b}$$

$$\widetilde{G}_{ABC} \equiv -\left[ \tfrac{1}{2} E_{[A} \widetilde{B}_{BC)} - \tfrac{1}{2} \check{C}_{[AB|}{}^D \widetilde{B}_{D|C)} \right] ~~. \tag{3.5c}$$

Any *tilded* superfield symbolizes the $i_\xi$-operation defined by $\widetilde{X}_{A_1 \cdots A_n} \equiv \xi^B X_{BA_1 \cdots A_n}$. The important point here is that even though the modified BIds (3.4a) and (3.4b) look different from the original ones (3.1), the formers are just rewriting of the latter. In other words, we can 'derive' (3.4a) and (3.4b) from (3.1), under the definition (3.5). In this sense, the $m$-modified system is equivalent to the original system (3.1), and therefore the same set of constraints (3.2) satisfies (3.4). This also solves the puzzle of 'free' parameter [4] for 11D supergravity which is supposed to be 'unique' excluding such parameters [15]. For this reason, we can use exactly the same constraint set (3.2) for our purpose from now.

---

[9]In an earlier version of this paper, there was a redundant $m \widetilde{G} \widetilde{B}$-term in the $\check{G}$-BId that should not be there.

[10]The difference between $\check{C}_{AB}{}^D$ and $C_{AB}{}^D$ does not matter in (3.5c), due to the identity $\xi^D \widetilde{B}_{DC} \equiv 0$. The overall negative sign in (3.5c) is caused by our universal definition of the *tilded* superfields, causing a flipping sign.

We now generalize this Abelian super Killing vector to a non-Abelian case that corresponds to a more general compactification $M_{11} \to M_D \times B$. According to the past experience of such gaugings in $\sigma$-models [16], the main change will be that the Lie derivative of the Killing vector no longer vanishes, but is proportional to the structure constant. Now such a super Killing vector is specified by the conditions

$$\mathcal{L}_{\xi^I} B_{MNP} \equiv \xi^{QI} \partial_Q B_{MNP} + \tfrac{1}{2}(\partial_{[M|} \xi^{QI}) B_{Q|NP)} \stackrel{*}{=} 0 \ , \tag{3.6a}$$

$$\mathcal{L}_{\xi^I} E_M{}^A \equiv \xi^{NI} \partial_N E_M{}^A + (\partial_M \xi^{NI}) E_N{}^A \stackrel{*}{=} 0 \ , \tag{3.6b}$$

$$\mathcal{L}_{\xi^I} \xi^{MJ} \equiv \xi^{PI} \partial_P \xi^{MJ} - \xi^{PJ} \partial_P \xi^{MI} \stackrel{*}{=} m^{-1} f^{IJK} \xi^{MK} \ , \tag{3.6c}$$

$$E_A \xi^{BI} \stackrel{*}{=} \xi^{CI} C_{CA}{}^B \ , \tag{3.6d}$$

$$\mathcal{L}_{\xi^I} C_{AB}{}^C = \xi^{DI} E_D C_{AB}{}^C \stackrel{*}{=} 0 \ . \tag{3.6e}$$

These are non-Abelian generalizations of (3.3).

We can try the non-Abelian generalization of the modified BIds (3.4). We encounter, however, an obstruction for the $\check{C}$-BId. This is because an $m^2$-term with the factor $\xi^{EI} \widetilde{B}_{EC}{}^J \equiv \xi^{EI} \xi^{FJ} B_{FEC} \neq 0$, no longer vanishes in those BIds due to the additional adjoint indices $I, J$, which used to vanish in the Abelian case.

Even though we do not yet have the solution to this problem, the important point here is that as long as we believe the uniqueness of the starting 11D superspace, along with the non-modified BIds, we still can formulate the non-Abelian minimal couplings in supermembrane in the next section. The main technical reason is that all we need for $\kappa$-invariance is the relationships like (3.6), with no need of modified superfield strengths.

## 4. Supermembrane with Non-Abelian Gauging

On the compactification of $M_{11} \to M_{10} \times S^1$ with Abelian gauging, we have seen in 11D superspace that all the new effects by the $m$-dependent terms cancel themselves. By the same token, the nontrivial-looking modified BIds turned out to be completely equivalent to conventional ones. This situation will be maintained for more generalized compactifications $M_{11} \to M_D \times B$. An intuitive explanation is that even though the original 11D will be compactified, the original superfield equations will be satisfied, and therefore, the original BIds will not be modified after all.

However, the effect of super Killing vectors corresponding to compactifications will have definitely non-trivial effects on a supermembrane action in 3d [8]. This is analogous to the gauging effect of any $\sigma$-models on $\mathcal{G}/\mathcal{H}$ with minimal couplings for the gauge subgroup $\mathcal{H}$ of $\mathcal{G}$ [16]. In particular, such minimal coupling can be introduced by a world-volume gauge field $A_i{}^I$.

These preliminaries at hand, we can give our supermembrane total action $I$ on 3d world-volume with the lagrangian:

$$I = \int d^3\sigma \left[ +\tfrac{1}{2}\sqrt{-g}\,g^{ij}\eta_{ab}\Pi_i{}^a\Pi_j{}^b - \tfrac{1}{2}\sqrt{-g} - \tfrac{1}{3}\epsilon^{ijk}\Pi_i{}^C\Pi_j{}^B\Pi_k{}^A B_{ABC} \right.$$
$$\left. + \tfrac{1}{2}m\epsilon^{ijk}\left(F_{ij}{}^I A_k{}^I - \tfrac{1}{3}f^{IJK}A_i{}^I A_j{}^J A_k{}^K\right) \right] \ . \tag{4.1}$$

We use the indices $i, j, \cdots = 0, 1, 2$ for the curved coordinates $(\sigma^i)$ of 3d world-volume, while $(Z^M) = (X^m, \theta^\mu)$ for the 11D superspace coordinates [6]. The $E_A{}^M$ is the vielbein in 11D superspace, and the pull-back $\Pi_i{}^A$ has non-Abelian minimal coupling

$$\Pi_i{}^A \equiv \left(\partial_i Z^M - mA_i{}^I \xi^{MI}\right) E_M{}^A \equiv \Pi_i^{(0)A} - mA_i{}^I \xi^{AI} \ , \tag{4.2}$$

with the coupling constant $m$. The original supermembrane action [8] can be recovered in the limit $m \to 0$. Needless to say, the Abelian case is also obtained as a special case by putting the structure constant to zero, with all related adjoint indices deleted. The $A_i{}^I = A_i{}^I(\sigma)$ is the non-Abelian gauge field on the world-volume with its field strength

$$F_{ij}{}^I \equiv \partial_i A_j{}^I - \partial_j A_i{}^I + f^{IJK} A_i{}^J A_j{}^K \ , \tag{4.3}$$

with the structure constant $f^{IJK}$ of the gauge group $G$. Even though (4.1) is for the non-Abelian gauging, any Abelian case can be also obtained by deleting all the adjoint indices $I, J, \cdots$.

As for the 11D superspace background, we adopt teleparallel superspace [13], for the same reason as the Abelian case. One intuitive reason is that it is more natural to use superspace constraints which do not have manifest local Lorentz covariance. One technical reason is that, as we will see, our action loses fermionic $\kappa$-invariance, when there is a Lorentz connection on the background superspace. For the reason already mentioned, we can use only the un-modified superfield strength $G_{ABCD}$ and $C_{AB}{}^C$ in teleparallel superspace formulation, instead of the $m$-modified ones.

Interestingly, since the $\pi_3$-homotopy mapping of a non-Abelian group is generally non-trivial, the constant $m$ in front of the Chern-Simons term is to be quantized, depending on $G$ for the compact space $B$. Specifically, $\pi_3(G) = \mathbb{Z}$ for $G = SO(n)\ (n \neq 4)$, $U(n)\ (n \geq$ 2), $SU(n)$ $(n \geq 2)$, $Sp(n)$ $(n \geq 1)$, $G_2$, $F_4$, $E_6$, $E_7$, or $E_8$, while $\pi_3(SO(4)) = \mathbb{Z} \oplus \mathbb{Z}$. For Abelian groups, such a mapping is trivial: $\pi_3(SO(2)) = \pi_3(U(1)) = 0$. For the group with $\pi_3(G) = \mathbb{Z}$, the quantization is [17]

$$m = \frac{n}{8\pi} \quad (n = \pm 1, \pm 2, \cdots) \ . \tag{4.4}$$

The local non-Abelian invariance of our action is given with the $\sigma$-dependent transformation parameter $\alpha^I$ by

$$\delta_\alpha A_i{}^I = +\partial_i \alpha^I + f^{IJK} A_i{}^J \alpha^K \equiv D_i \alpha^I \ , \tag{4.5a}$$

$$\delta_\alpha Z^M = +m\,\alpha^I\,\xi^{MI} \ , \tag{4.5b}$$

$$\delta_\alpha \xi^{MI} = +m\alpha^J\,\xi^{NJ}\partial_N \xi^{MI} \ , \quad \delta_\alpha \xi^A \stackrel{*}{=} -f^{IJK}\alpha^J \xi^{AK} \ , \tag{4.5c}$$

$$\delta_\alpha E_M{}^A \stackrel{*}{=} -m\,\alpha^I\,(\partial_M \xi^{NI})E_N{}^A \ , \quad \delta_\alpha E_A{}^M \stackrel{*}{=} +m\,\alpha^I\,E_A \xi^{MI} \ , \tag{4.5d}$$

$$\delta_\alpha \Pi_i{}^M = +m\,\alpha^I\,\Pi_i{}^N \partial_N \xi^{MI} \ , \quad \delta_\alpha \Pi_i{}^A = 0 \ , \tag{4.5e}$$

$$\delta_\alpha F_{ij} = 0 \ , \tag{4.5f}$$

$$\delta_\alpha B_{MNP} \stackrel{*}{=} -\tfrac{1}{2}m\alpha^I (\partial_{[M|}\xi^{QI})B_{Q|NP)} \ , \quad \delta_\alpha B_{ABC} = 0 \ . \tag{4.5g}$$

An Abelian case is easily obtained by the truncation of the adjoint indices and the structure constant. All the (super)fields carrying curved 11D superspace indices transform non-trivially, *except for* $A_i{}^I$. The local invariance of $\delta_\alpha I = 0$ under $G$ is easily confirmed, because of the invariances of $\Pi_i{}^A$ and $F_{ij}$.

Our action is also invariant under the $\Lambda$-gauge transformation rule

$$\delta_\Lambda B_{ABC} = +\tfrac{1}{2}E_{[A}\Lambda_{BC)} - \tfrac{1}{2}C_{[AB|}{}^D \Lambda_{D|C)} \ , \quad \xi^{AI}E_A \Lambda_{BC} \stackrel{*}{=} 0 \ , \tag{4.6a}$$

$$\delta_\Lambda E_A{}^M = +m\widetilde{\Lambda}_A{}^I \xi^{MI} \ , \quad \delta_\Lambda E_M{}^A = -m\widetilde{\Lambda}_M{}^I \xi^{AI} \ , \tag{4.6b}$$

$$\delta_\Lambda A_i{}^I = -\Pi_i{}^A \widetilde{\Lambda}_A{}^I \equiv -\widetilde{\Lambda}_i{}^I \ , \quad \widetilde{\Lambda}_A{}^I \equiv \xi^{BI}\Lambda_{BA} \ , \tag{4.6c}$$

$$\delta_\Lambda \Pi_i{}^A = 0 \ , \quad \delta_\Lambda g_{ij} = 0 \ , \quad \delta_\Lambda Z^M = 0 \ , \quad \delta_\Lambda \xi^{AI} = 0 \ , \quad \delta_\Lambda \xi^{MI} = 0 \ . \tag{4.6d}$$

We have $\delta_\Lambda \Pi_i{}^A = 0$, justifying the minimal coupling in $\Pi_i{}^A$. We easily see that the crucial $F_{ij}$-linear terms in $\delta_\Lambda I$ will be cancelled by the variation of the Chern-Simons term.

We now study the fermionic $\kappa$-invariance [9][8]. Our action $I$ is invariant under

$$\delta_\kappa E^\alpha \equiv (\delta_\kappa Z^M)E_M{}^\alpha = (I + \Gamma)^{\alpha\beta}\kappa_\beta \equiv [(I+\Gamma)\kappa]^\alpha \ , \tag{4.7a}$$

$$\delta_\kappa E^a \equiv (\delta_\kappa Z^M)E_M{}^a = 0 \ , \quad \Gamma \equiv +\tfrac{i}{6\sqrt{-g}}\epsilon^{ijk}\Pi_i{}^a \Pi_j{}^b \Pi_k{}^c \gamma_{abc} \ , \tag{4.7b}$$

$$\delta_\kappa A_i{}^I = \Pi_i{}^A \xi^{BI}(\delta_\kappa E^C)B_{CBA} \equiv \Pi_i{}^A \xi^{BI} \Xi_{BA} \equiv \Pi_i{}^A \widetilde{\Xi}_A{}^I \ , \quad \Xi_{AB} \equiv (\delta_\kappa E^C)B_{CAB} \ , \tag{4.7c}$$

$$\delta_\kappa E_A{}^M = (\delta_\kappa E^B) E_B E_A{}^M - m\widetilde{\Xi}_A{}^I \xi^{MI} \ , \quad \delta_\kappa E_M{}^A = (\delta_\kappa E^B) E_B E_M{}^A + m\widetilde{\Xi}_M{}^I \xi^{AI} \ , \quad (4.7\text{d})$$

$$\delta_\kappa \xi^{AI} = (\delta_\kappa E^C) \xi^{BI} C_{BC}{}^A \ , \quad (4.7\text{e})$$

$$\delta_\kappa \Pi_i{}^A = \partial_i (\delta_\kappa E^A) + (\delta_\kappa E^C) \Pi_i{}^B C_{BC}{}^A \ , \quad (4.7\text{f})$$

$$\delta_\kappa B_{ABC} = (\delta_\kappa E^D) E_D B_{ABC} \ . \quad (4.7\text{g})$$

As stated in [13], (4.7f) takes a simpler form than Lorentz covariant formulation [7]. Needless to say, $\Pi_i{}^A$ in this equation contains the $m$-term, but still no $m$-explicit term arises in (4.7f). As is easily seen, the $m$-dependent terms in (4.7d) and $\delta_\kappa A_i$ itself are the special cases of the $\Lambda$-transformation rules (4.6b) and (4.6c) with $\Lambda_{AB} \equiv -\Xi_{AB} \equiv -(\delta_k E^C) B_{CAB}$. Note, however, $\delta_\kappa B_{ABC}$ has *no* corresponding term. This is because otherwise all of them cancel each other due to $\delta_\Lambda I = 0$. The effect of having the $\Xi$-terms only for $\delta_\kappa E_M{}^A$, $\delta_\kappa E_A{}^M$ and $\delta_\kappa A_i$ is to cancel unwanted terms in $\delta_\kappa I$ arising otherwise.

The $\kappa$-invariance of our action can be confirmed in a way parallel to the original super-membrane case [8], with subtle differences by the $m$-dependence and non-Abelian feature of super Killing vectors. The algebraic $g_{ij}$-field equation takes exactly the same form as the embedding condition in the conventional case [8]:

$$g_{ij} \doteq \Pi_i{}^a \Pi_{ja} \ , \quad (4.8)$$

where $\doteq$ is for a field equation. Needless to say, our pull-backs contain also the $m$-dependent terms. Other relationships involving $\Gamma$ are exactly same as the conventional case [8] or the Abelian case:

$$\Gamma^2 \doteq +I \ , \quad \epsilon^{ijk} \gamma_{jk} \Gamma \doteq -2i\sqrt{-g}\, \gamma_i \ ,$$
$$\gamma_i \equiv +\Pi_i{}^a \gamma_a \ , \quad \gamma_{ij} \equiv \Pi_i{}^a \Pi_j{}^b \gamma_{ab} \ . \quad (4.9)$$

As in the Abelian gauging, the confirmation $\delta_\kappa I = 0$ needs also important relationships, such as

$$\partial_{[i} \Pi_{j]}{}^A = \Pi_i{}^B \Pi_j{}^C C_{CB}{}^A - m F_{ij}{}^I \xi^{AI} \ , \quad (4.10\text{a})$$

$$\mathcal{L}_\xi B_{ABC} = \xi^{DI} E_D B_{ABC} \stackrel{*}{=} 0 \ . \quad (4.10\text{b})$$

The latter is confirmed by (3.6a), while (4.10a) needs the relationship

$$m \xi^{BI} \xi^{CJ} C_{CB}{}^A \stackrel{*}{=} f^{IJK} \xi^{AK} \ , \quad (4.11)$$

derived from (3.6c). An Abelian gauging can be also obtained by truncating the adjoint indices and structure constants.

One of the most crucial cancellation in the action invariance $\delta_\kappa I = 0$ arises out of the Wess-Zumino-Witten term: (i) From the partial integration of $\partial_i$ in $\epsilon^{ijk}[\partial_i(\delta_\kappa E^C)]\Pi_j{}^B\Pi_k{}^A B_{ABC}$ hitting $\Pi_j{}^B$ producing a term with $mF_{ij}$. (ii) From the variation $\delta_\kappa A_i$ in the Chern-Simons term, yielding a term with $m\epsilon^{ijk}\widetilde{\Xi}_i F_{jk}$. Both of these have the same structure cancelling each other. This cancellation also justifies the necessity of the constant $m$ in the Chern-Simons term, which is the minimal coupling constant at the same time.

As we have seen, it is not only the $\Lambda$-invariance, but also the $\kappa$-invariance that necessitates the Chern-Simons term. There are other reasons that we should have the Chern-Simons term. For example, if there were no Chern-Simons term, the minimal couplings of $A_i$ to the superspace coordinates $Z^M$ or $g_{ij}$ would result in additional constraints, spoiling the original physical degrees of freedom of these fields. Thanks to our Chern-Simons term, such constraints will not arise, but all the minimal coupling terms contribute only as the source term $J^i$ to the vector field equation as $\epsilon^{ijk}F_{jk}{}^I \doteq J^{iI}$. This also makes the whole system nontrivial, because our newly-introduced gauge field couples to conventional fields $Z^M$ in a nontrivial way, still respecting their original degrees of freedom.

We have been using teleparallel superspace as the consistent background for our supermembrane modified by the super Killing vector $\xi^A$. The most important technical reason is the problem with conventional constraints for the $\kappa$-invariance of our action that should be addressed here. Suppose we adopt Lorentz covariant formulation, replacing (3.3d) and (4.10a) now by

$$\nabla_A \xi^{BI} \stackrel{*}{=} \xi^{CI} T_{CA}{}^B \ , \tag{4.12a}$$

$$\nabla_{[i}\Pi_{j]}{}^A = \Pi_i{}^B \Pi_j{}^C T_{CB}{}^A - mF_{ij}{}^I \xi^{AI} + mA_{[i}{}^I \Pi_{j]}{}^C \xi^{BI} \omega_{BC}{}^A \ , \tag{4.12b}$$

where $\nabla_i$ is a Lorentz covariant derivative acting like $\nabla_i X_A \equiv \partial_i X_A + \Pi_i{}^A \omega_{AB}{}^C X_C$. Note that the last term in (4.12b) arises from the difference between $\Pi_{[i|}^{(0)B} \omega_B{}^{AC} \Pi_{|j]C}$ and $\Pi_{[i|}{}^B \omega_B{}^{AC} \Pi_{|j]C}$. Now the problem is that when we vary our action under $\delta_\kappa$, the Wess-Zumino-Witten term yields an additional term proportional to $m\epsilon^{ijk}\Pi_i{}^C A_j \Pi_k{}^D \xi^F \omega_{FD}{}^B (\delta_\kappa E^E) B_{EBC}$ that has no other counter-terms to cancel. On the other hand, teleparallel superspace has *no* such an $\omega$-dependent term generated, thanks to the absence of manifest local Lorentz covariance from the outset.

As far as the target 11D superspace is concerned, there is no physical difference between teleparallel superspace [13] and the conventional superspace [7]. However, when it comes to the physics of supermembrane on 3d, we have seen such a great difference due to the valid fermionic $\kappa$-invariance of the action. This seems to tell us that only teleparallel superspace [13] with no manifest local Lorentz covariance, is the most suitable and consistent with the super Killing vector introduced for the compactification from 11D into 10D. Since supermembrane is an important 'probe' of superspace background, our result indicates the importance of teleparallel superspace for compactifications of 11D or M-theory itself.

Before concluding this section, we give here all the field equations of our fields $g_{ij}$, $Z^M$ and $A_i{}^I$ in 3d, as

$$g_{ij} \doteq \Pi_i{}^a \Pi_{ia} ~, \tag{4.8}$$

$$\delta_A{}^a \partial_i(\sqrt{-g}\, \Pi^i{}_a) - \sqrt{-g}\, \Pi_i{}^B C_{BA}{}^d \Pi^i{}_d$$
$$\doteq +\tfrac{1}{3}\epsilon^{Ijk}\Pi_i{}^D\Pi_j{}^C\Pi_k{}^B G_{BCDA} - m\epsilon^{ijk} F_{ij}{}^I \xi^{BI}\Pi_k{}^C B_{CBA} ~, \tag{4.13a}$$

$$\epsilon^{ijk}(F_{jk}{}^I - \widetilde{B}_{jk}) \doteq \sqrt{-g}\,\Pi^{ia}\xi_a{}^I ~. \tag{4.13b}$$

Compared with the original supermembrane case [8], we have the $A$-field equation as an extra, while the super Killing vector containing terms are the new effects here. All other terms are formally the same as the $m=0$ case.

The mutual consistency between (4.13a) and (4.13b) can be confirmed by taking the divergence of the latter. In fact, we get

$$\begin{aligned}
0 &\stackrel{?}{=} D_i\big(\epsilon^{ijk} F_{jk}{}^I - \sqrt{-g}\,\Pi^{ia}\xi_a{}^I + \epsilon^{ijk}\xi^{AI}\Pi_j{}^C\Pi_k{}^B B_{BCA}\big) \\
&= -[\partial_i(\sqrt{-g}\,\Pi^{ia})]\xi_a{}^I - \sqrt{-g}\,\Pi^{ia}\Pi_i{}^B \xi^{EI} C_{EB}{}^a \\
&\quad - m\epsilon^{ijk}\xi^{AI} F_{ij}{}^J \xi^{BJ}\Pi_k{}^B B_{BCA} + \tfrac{1}{3}\epsilon^{ijk}\xi^{AI}\Pi_j{}^C\Pi_k{}^B\Pi_i{}^D G_{DBCA} \\
&\doteq 0 ~. 
\end{aligned} \tag{4.14}$$

This vanishes, because the penultimate side is nothing but the multiplication of the $Z^M$-field equation (4.13a) by $\xi^{AI}$. Use has been also made of the relation (4.10b).

## 5. Concluding Remarks

In this paper, we have performed the non-Abelian gauging of supermembrane, by introducing a vector field on its world-volume. We have confirmed that our action has three invariances under fermionic $\kappa$-symmetry, local non-Abelian gauge symmetry, and composite $\Lambda$-symmetry for the antisymmetric tensor $B_{ABC}$. We have seen that the $\Lambda$-invariance requires the minimal couplings to the super Killing vector $\xi^{AI}$, while both $\Lambda$- and $\kappa$-invariances necessitate the Chern-Simons term, which makes our system both consistent and nontrivial.

Since the $\pi_3$-mapping of a non-Abelian gauge group $G$ associated with the compact space $B$ is generally non-trivial, the $m$-coefficient of our Chern-Simons term is to be quantized. This situation is different from an Abelian gauging with $\pi_3(U(1)) = 0$. Even though the precise significance of this quantization is yet to be studied, we stress that it is our formulation that revealed such a quantization in terms of supermembrane action principle for 3d physics.

The Abelian gauging requires a vector field on the world-volume, which is similar to the Abelian vector used in D-branes [14]. Even though we do not yet know any direct relationships, it is quite natural to have the D-brane generalization of our formulation.

Our results in this paper tell two important aspects for M-theory. First, the introduction of a super Killing vector $\xi^{AI}$ with the parameter $m$ seems to induce *no* new physical effects on the target 11D superspace itself, because all the field strengths and BIds are entirely reduced to the original ones at $m = 0$ in 11D [7]. This is also consistent with our past experience, *i.e.*, any naïve modification of 11D supergravity [3] is bound to fail, due to the 'uniqueness' of 11D supergravity [15], unless it is related to certain M-theory higher-order correction terms. Second, most importantly, the existence of the super Killing vector $\xi^{AI}$ induces nontrivial physical effects on the supermembrane action in 3d, despite no seeming physical effects on the 11D target superspace. The quantization of the Chern-Simons term also support the non-trivial feature of the system on the world-volume. To put it differently, while 11D supergravity is 'unique' [3][7], there are still some ambiguities for supermembrane physics in the 3d world-volume. Our results have uncovered such nontrivial unknown aspects of double-compactifications of M-theory.

To our knowledge, our formulation is the first one that provides with the non-Abelian minimal couplings into supermembrane action in 11D with a Chern-Simons term. These nontrivial couplings make double-compactifications [1] more interesting, because without supermembrane action on 3d, all the effects of the super Killing vector $\xi^{AI}$ simply disappeared within 11D target superspace. It is these non-Abelian couplings that make the new effects of $\xi^{AI}$ more nontrivial, interacting with physical fields in supermembrane action. Additionally, our non-Abelian gauge field is neither auxiliary nor composite as in the past references [4], but is 'topological' with a proper Chern-Simons term. Since supermembrane [8] is an important 'probe' for 11D backgrounds, our result indicates important effects of super Killing vector for compactifications on the supermembrane world-volume physics.

The techniques developed in this paper will play an important role, when considering general compactifications of M-theory, such as compactifications into superstring in 10D or lower-dimensions. This is because these techniques are based on the supermembrane world-volume physics, instead of exact solutions or the end results after compactifications. As an important probe of superspace backgrounds including both the compactified and the original superspaces, our formulation will be of great importance, with all the effects of compactifications crystalized with the couplings to super Killing vectors.

In this paper we have seen that teleparallel superspace is the consistent background for supermembranes with the non-Abelian super Killing vector. We have seen this with the failure of $\kappa$-invariance of the action in conventional Lorentz covariant superspace. This result is also natural from the viewpoint that compactifications such as that from 11D into 10D necessarily break local Lorentz symmetry within 11D. Supermembrane physics, as an important probe for 11D background, has revealed the significance of teleparallel superspace in the compactification of 11D superspace or even M-theory itself. We emphasize that it is no longer just for curiosity that we study teleparallel superspace [13], but it is also based on fundamental significance related to supermembrane physics in 3d [8]. In this sense, teleparallel superspace [13] is more than just 'a technical tool', but a consistent (probably unique) background, when considering the double-compactification [1] of supermembrane [8]. The importance of teleparallel formulation with no manifest local Lorentz symmetry [13] should be re-stressed in the context of double-compactifications [1] of supermembranes [8].

We expect more interesting results to be developed in these new directions. We are grateful to W. Siegel for discussing $\kappa$-symmetry.